\documentclass[11pt]{article}
\usepackage{graphicx}
\usepackage{float} 
\usepackage{amsmath}
\usepackage{amsfonts}   
\usepackage{amssymb}

\def\l{\lambda}

\def\s{\sigma} 

\def\th{\theta}
\def\ph{\phi} 
\def\m{\mu}

\def\r{\ref}
\def\p{\partial}
\def\no{\nonumber}
\def\f{\frac}
\def\S{\Sigma}
\def\D{\Delta}
\def\Om{\Omega}
\def\c{\cite}
\begin{document}
\date{}
\title{{\bf{\Large Killing Symmetries and Smarr formula for Black Holes in Arbitrary Dimensions}}}
\author{
{\bf {{\normalsize Rabin Banerjee}}$
$\thanks{e-mail: rabin@bose.res.in}}
,$~${\bf {\normalsize Bibhas Ranjan Majhi}$
$\thanks{e-mail: bibhas@bose.res.in}}
,$~${\bf {\normalsize Sujoy Kumar Modak}$
$\thanks{e-mail: sujoy@bose.res.in}}\\
 {\normalsize S.~N.~Bose National Centre for Basic Sciences,}
\\{\normalsize JD Block, Sector III, Salt Lake, Kolkata-700098, India}
\\[0.3cm]and\\[0.3cm]
{\bf {\normalsize Saurav Samanta}$
$\thanks{e-mail: srvsmnt@gmail.com}}\\
 {\normalsize Narasinha Dutt College}
\\{\normalsize 129, Belilious Road, Howrah-711101, India}
\\[0.3cm]
}

\maketitle


\begin{abstract}
We calculate the effective Komar conserved quantities for the $N+1$ dimensional charged Myers-Perry spacetime. At the event horizon we derive a new identity $K_{\chi^{\mu}}=2ST$ where the left hand side is the Komar conserved quantity corresponding to the null Killing vector $\chi^{\mu}$ while in the right hand side $S,~T$ are the black hole entropy and Hawking temperature. From this identity we also derive the generalized Smarr formula connecting the macroscopic parameters $M,~J,~Q$ of the black hole with its surface gravity and horizon area. The consistency of this new formula
is established by an independent algebraic approach.
\end{abstract}

\section{Introduction}
Einstein's general theory of relativity describes gravity as a 3+1 dimensional curved spacetime. A remarkable prediction of this theory is the existance of black holes which represent disconnected regions of spacetime manifold. The event horizon of a black hole separates it from the rest of the spacetime. This horizon acts as a one way membrane so that particles can enter in a black hole but nothing can escape from it. According to the no hair theorem, the most general black hole solution (Kerr-Newman) in 3+1 dimensions can have only three parameters; mass, charge and  angular momentum. In 1973, Smarr \c{smarr} showed that the mass of a Kerr-Newman black hole can be written as a sum of three terms; surface energy, rotational energy and electromagnetic energy. The infinitesimal surface energy which consists of a product between surface gravity and increment of horizon area was identified as entropic energy by Bekenstein \cite{Beken1,Beken2,Beken3}. This idea was put on a firm footing when Hawking \cite{Hawking1,Hawking2} showed that, contrary to the classical picture, black holes radiate energy and have temperature and entropy.

The idea of higher dimension, which at first appeared as a mathematical generalization, has some interesting applications in physics. Considering compactification of extra dimensions, it has been suggested that, the large hierarchy between the weak scale and the fundamental scale of gravity can be eliminated \c{lisa1, lisa2}. In such a scenario black holes may appear as higher dimensional objects if the horizon area of a black hole is much smaller than the size scale of the extra dimensions \c{aliev}. As an approximation one can describe these black holes as solutions of higher dimensional Einstein's theory. In fact, it has been even argued that there is a possibility of creation of higher dimensional black holes in future Large Hadron Collider \cite{steven}. From this perspective studying black hole physics in arbitrary dimension has some theoretical as well as phenomenological importance. Though the spherically symmetric black hole solutions of higher dimensional Einstein's theory was discovered long ago \c{tangher}, research in this area remained dormant for quite sometime. However, observing the applications of higher dimensional theories, interest was renewed and different other black hole solutions were found \c{myers,dianyan,aliev}. Among these, the spherically symmetric solutions are unique solutions of Einstein's equation in higher dimensions but the rotating black holes are not \c{emparan1}. In fact in (4+1) dimensions a black ring solution appears together with the rotating solution having the same mass and angular momentum \c{emparan2}{\footnote{for an extensive review on higher dimensional black holes see \c{emparan3}}}. Though the properties of rotating black holes \c{myers} were studied thoroughly but the generalization to the charged case\c{dianyan,aliev} has not recieved much attention. 

In the present paper we perform a detailed study to explore various important features of charged, rotating black holes in arbitrary dimensions. We first discuss the thermodynamical properties of charged Myers-Perry black holes. Then using an algebraic approach, motivated by the work of Smarr \c{smarr}, we find a mass formula for $N+1$ dimensional Reissner-Nordstrom black hole. However this method cannot be generalized to the rotating case. We therefore develop an alternative scheme which involves the knowledge of conserved quantities of the respective spacetime. These conserved quantities are related to various Killing symmetries and are found by evaluating the Komar integrals. The Komar integrals are evaluated at the boundary of a spatial hypersurface in a spacetime. The exact nature of this surface must be $r=$ constant with time synchronised events \cite{cohen1,cohen2,modak}. This gives freedom to calculate these conserved quantities explicitly on any such surface, which need not be the asymptotic surface or event horizon of a black hole. These quantities are used to find the Komar conserved quantity ($K_{\chi^{\mu}}$) at the event horizon corresponding to the null Killing vector $\chi^{\mu}$. This leads to the remarkable dimension independent identity, $K_{\chi^{\mu}}=2ST$. Here ($S$) and $T$ are the black hole entropy and Hawking temperature respectively. The profound nature of this identity is that although each individual term is dimension dependent they together satisfy an identity which is completely dimension independent. As an application of this identity we derive the generalised Smarr formula for the charged Myers-Perry black holes.

To put our analysis in a proper perspective let us compare it with other approaches for obtaining the Smarr formula.  Before that let us recall that our obtention of the Smarr formula is a byproduct of the identity $K_{\chi^{\mu}}=2ST$ elaborated above. In our analysis the emphasis is more on this identity and on the derivation of various Komar integrals leading to it. The other approaches that we now discuss are basically geared for only deriving the Smarr formula. The original method of Smarr \c{smarr}, applicable to four dimensions, was based on exploiting Euler's theorem on homogeneous functions. We show in section 3 that, while it is possible to extend the method to arbitrary dimensional Reissner-Nordstrom black holes, the same is not obvious for the charged Myers-Perry example. The other approach is due to Bardeen, Carter and  Hawking \c{Bardeen} which is metric independent. But here also the results are available in four dimensions. The derivation of Smarr formula for higher dimensional Kerr-AdS black holes has been recently provided by Barnich and Compere \c{barnich}. In this method they modify the surface integrals to compute various conserved charges for both flat and AdS backgrounds. One can also try to generalise this approach for the charged Myers-Perry spacetime where the solution space is  actually parametrized by three black hole parameters $M,~a,~Q$. In this paper, however, we are adopting a different approach to derive the generalised Smarr formula for this spacetime. We do not modify the surface integrals. Rather, we evaluate the standard Komar integrals at the boundary of a $N$ dimensional spatial hypersurface. This boundary is at constant finite radial distance and contains only time synchronised events. As a result we are able to explicitly derive the effective values of Komar conserved quantities. At asymptotic infinity they reproduce the observed mass($M$) and angular momentum($J$) of the black hole.   Using the Komar conserved quantities we derive the identity $K_{\chi_{\mu}}=2ST$. The generalised Smarr formula follows from this identity. This formula is also shown to be compatible with the first law of thermodynamics.

The organization of the paper is as follows. In the second section, we study the thermodynamical properties of the charged Myers-Perry black hole in arbitrary dimensions. Using an algebraic approach we derive the Smarr formula for the $N+1$ dimensional Reissner-Nordstrom black holes in section 3. After that in section 4 we emphasize the importance of Killing symmetries to find the effective Komar conserved quantities. At the event horizon, these results are used to obtain the identity $K_{\chi^{\mu}}=2ST$. This identity is used in section 5 to find the generalized Smarr formula and first law of black hole thermodynamics. We conclude in section 6. There are two appendices; dimensional reduction of the metric is discussed in appendix A while Komar integrands are explicitly given in appendix B.

\section{The higher dimensional black holes}

The spacetime metric for the $N+1$ dimensional charged Myers-Perry black hole in Boyer-Lindquist type coordinates, with one spin parameter ($a$), is given by \c{aliev,dianyan}
\begin{eqnarray}
ds^2&=&-\left(1-\f{m}{r^{N-4}\Sigma}+\frac{q^2}{r^{2(N-3)}\Sigma}\right)dt^2+\f{r^{N-2}\Sigma}{\D}dr^2+\S d\th^2-\f{2a(mr^{N-2}-q^2)\sin^2\th}{r^{2(N-3)}\S}dtd\ph\no\\
&&+\left(r^2+a^2+\f{a^2(mr^{N-2}-q^2)\sin^2\th}{r^{2(N-3)}\S}\right)\sin^2\th d\ph^2+r^2\cos^2\th d\Om^2_{N-3},
\label{metric}
\end{eqnarray}
with the following identifications,
\begin{eqnarray}
\D&=&r^{N-2}(r^2+a^2)-mr^2+q^2r^{4-N},
\label{delta}\\
\S&=&r^2+a^2\cos^2\th,
\label{sigma}
\end{eqnarray}
\begin{eqnarray}
d\Om^2_{N-3}=d\chi_1^2+\sin^2\chi_1[d\chi^2_2+\sin^2\chi_2(\cdot\cdot\cdot d\chi^2_{N-3})].
\label{omega}
\end{eqnarray}
The electromagnetic potential one form for the spacetime (\r{metric}) is 
\begin{eqnarray}
A=A_{\mu}dx^{\mu}=-\f{Q}{(N-2)r^{N-4}\Sigma}(dt-a\sin^2\theta d\phi).
\label{phi}
\end{eqnarray}
In appropriate limits this metric reproduces the $N+1$ dimensional spherically symmetric, static Schwarzscild, Reissner-Nordstrom \c{tangher} and axially symmetric, rotational Myers-Perry spcetime \c{myers}.

It must be stated that the metric (\ref{metric}) is a solution of the Einstein-Maxwell equation only for linear order in $a$ \c{aliev}. However   in our subesequent analysis we keep all terms involving $a$. Finally it will be shown that the arbitrary dimensional result for the Smarr formula, compatible with black hole thermodynamics, can be recovered only if terms linear in $a$ are retained.

The determinant ($g$) of the metric (\r{metric}) gives
\begin{eqnarray}
\sqrt{-g}=\sqrt{\gamma}\S r^{N-3}\sin\th\cos^{N-3}\th,
\label{deter}
\end{eqnarray}
where $\gamma$ is the determinant of the metric (\r{omega}). The parametrers $m,~a,~q$ are related with the physical mass ($M$), angular momentum ($J$) and charge ($Q$) through the relations given by
\begin{eqnarray}
M &=&\f{A_{N-1}(N-1)}{16\pi G}m
\label{m}\\
J &=&\f{A_{N-1}}{8\pi G}ma
\label{j}\\
Q &=&\pm\sqrt{\f{(N-2)(N-1)A_{N-1}}{8\pi G}}q
\label{q}
\end{eqnarray}
Here $A_{N-1}$ is the area of the unit sphere in $N-1$ dimensions 
\begin{eqnarray}
A_{N-1}=\int_{0}^{2\pi}d\ph\int_0^\pi \sin\th\cos^{N-3}\th d\th\huge\prod_{i=1}^{N-3}\int_{0}^{\pi}\sin^{(N-3)-i}\chi_{i}d\chi_{i}=\f{2\pi^{N/2}}{\Gamma(N/2)}.
\label{a}
\end{eqnarray}
The position of the event horizon is represented by the largest root of the polynomial $\Delta_{r=r_+}=0$. The angular velocity at the event horizon is given by
\begin{eqnarray}
\Omega_{H}=\f{a}{r_+^2+a^2}.
\label{ang}
\end{eqnarray}

Let us now calculate the area of the event horizon for the spacetime metric (\r{metric}). This is given by the standard definition of horizon area by integrating the positive square-root of the induced metric ($\eta$) on the $N-1$ dimensional spatial angular hypersurface ($\Sigma$) for fixed $r=r_+$  
\begin{eqnarray}
A_{H}=\int_{\Sigma(r=r_+)}\sqrt{|\eta|}~d^{N-1}\Sigma .
\label{hora}
\end{eqnarray}
The result of the integration yields,
\begin{eqnarray}
A_{H}=\f{4\pi r_+^{N-3}(r_+^2+a^2)\bar{A}_{N-3}}{N-2}.
\label{hora1}
\end{eqnarray}
Here $\bar{A}_{N-3}$ is the area of $N-3$ dimensional angular hypersurface represented by the angular variables $\chi_j$-s
\begin{eqnarray}
\bar{A}_{N-3}=\huge\prod_{i=1}^{N-3}\int_{0}^{\pi}\sin^{(N-3)-i}\chi_{i}d\chi_{i}=\frac{\pi^{\frac{N}{2}-1}}{\Gamma(\frac{N}{2}-1)}.
\label{an-3}
\end{eqnarray}
Therefore the semiclassical black hole entropy, as given by the Bekenstein-Hawking formula, yields 
\begin{eqnarray}
S=\f{A_H}{4}=\f{\pi r_+^{N-3}(r_+^2+a^2)\bar{A}_{N-3}}{N-2}.
\label{entr}
\end{eqnarray}

     To find the Hawking temperature we shall take help of Hawking's periodicity argument \c{Hawking3}. Such analysis can be simplified by considering the near horizon effective metric of (\ref{metric}). Near the event horizon the effective theory is driven by the two dimensional ($t-r$) metric \cite{carlip}-\cite{kumet}. In our case the two dimensional metric is given by (see appendix A) (\r{effective1}). Now the metric (\ref{effective1}) has singularity at $r=r_+$. To remove this singularity consider the following transformation
\begin{eqnarray}
x = \frac{1}{\kappa} F^{1/2}(r)
\label{trans}
\end{eqnarray}
where $\kappa$, the surface gravity of the black hole (\ref{metric}), is given by
\begin{eqnarray}
\kappa = \frac{F'(r_+)}{2}=\f{(N-4)(r_+^2+a^2)+2r_+^2-(N-2)q^2r_+^{2(3-N)}}{2 r_+(r_+^2+a^2)}.
\label{surface}
\end{eqnarray}
Substituting (\ref{trans}) in (\ref{effective1}) and then euclideanizing (i.e. $t=i\tau$)  we obtain,
\begin{eqnarray}
ds^2_{eff} = (\kappa x)^2 d\tau^2 + \Big(\frac{2\kappa}{F'(r)}\Big)^2dx^2.
\label{effective3}
\end{eqnarray}
This metric is regular at $x = 0$ and $\tau$ is regarded as an angular variable with period $2\pi/\kappa$, which is regarded as the inverse of Hawking temperature (in units of $\hbar = 1$). Hence the explicit expression for Hawking temperature is given by,
\begin{eqnarray}
T= \frac{\kappa}{2\pi} = \f{(N-4)(r_+^2+a^2)+2r_+^2-(N-2)q^2r_+^{2(3-N)}}{4\pi r_+(r_+^2+a^2)}.
\label{temp}
\end{eqnarray}
In appropriate limits, (\r{entr}) and (\r{temp}) reproduce the entropy and Hawking temperature for all other black holes in $N+1$ dimensions. 


\section{Algebraic approach to Smarr formula in Reissner-Nordstrom black holes}
In the last section we have introduced the higher dimensional black holes in Einstein and Einstein-Maxwell gravity. In this section we shall derive a mass formula for $N+1$ dimensional Reissner-Nordstrom black holes. This will be done in an algebraic approach which is a generalisation of the method used by Smarr \cite{smarr}. 

The metric for the $N+1$ dimensional R-N black hole is easily found by setting $a=0$ in (\r{metric}). Similarly all other physical entities like the horizon radius, horizon area, Hawking temperature and black hole entropy for this spacetime are obtained from the respective expressions of section1. 

The horizon condition, $\Delta|_{r=r_+}=0$ (for $a=0$), is simplified as
\begin{eqnarray}
r_+^{N-2}=\frac{m}{2}+\sqrt{\frac{m^2}{4}-q^2}
\label{rnhor}
\end{eqnarray}
Substituting this result in the expression of horizon area (\ref{hora1}) (with $a=0$), we find 
\begin{eqnarray}
A_{H}=\frac{4\pi \bar{A}_{N-3}}{N-2}\left(\frac{m}{2}+\sqrt{\frac{m^2}{4}-q^2}\right)^{\f{N-1}{N-2}}.
\label{rnarea}
\end{eqnarray}
Now inverting this relation we obtain $m$ in terms of $A_{H}$ and $q$, given by
\begin{eqnarray}
m=\left(\frac{(N-2)A_{H}}{4\pi \bar{A}_{N-3}}\right)^{\frac{N-2}{N-1}}+q^2\left(\frac{4\pi \bar{A}_{N-3}}{(N-2)A_{H}}\right)^{\frac{N-2}{N-1}}.
\label{rnmass}
\end{eqnarray} 
The differential form of this relation yields
\begin{eqnarray}
dm &=&t~dA_{H}+\tilde\ph ~dq\label{rnmadss}
\end{eqnarray}
where,
\begin{eqnarray}
t &=& \f{\p m}{\p A_H} = \f{(N-2)}{(N-1)\bar{A}_{N-3}}T\label{t}\\
\tilde\ph &=& \f{\p m}{\p q} = \f{2q}{(N-2)r_+^{N-2}}\label{p1}
\end{eqnarray}
and the Hawking temperature $T$ is defined in (\r{temp}) with $a=0$. From (\r{rnmass}) we find that $m$ is a homogeneous function of degree $\frac{N-2}{N-1}$ in ($A_{H}$, $q^\f{N-1}{N-2}$) respectively. Therefore making use of the Euler's theorem on homogeneous functions it follows that
\begin{eqnarray}
m=\frac{(N-1)}{(N-2)}~tA_{H}+\tilde\ph~ q.
\label{rnsmarr1}
\end{eqnarray}
Now let us express this relation in terms of physical mass ($M$), charge ($Q$), temperature ($T$) and electric potential ($\Phi$) by using (\r{m}), (\r{q}), (\r{temp}) and (\r{phi}) respectively. Here $\Phi$ is the timelike part of the guage potential $A_{\mu}=-\f{Q}{(N-2)r^{N-2}}(1,0,0,0,\cdot\cdot\cdot)$ (also follows from (\r{phi}) with $a=0$) and this can be calculated as follows. The $N+1$ dimensional R-N black hole being spherically symmetric, the global timelike Killing vector is $\xi^{\mu}=(1,0,0,0,\cdot\cdot\cdot)$. Therefore the scalar potential is found to be
\begin{eqnarray}
\Phi=\xi^{\mu}A_{\mu}\big|_{r=\infty}-\xi^{\mu}A_{\mu}\big|_{r=r_+}=0-\left(-\f{Q}{(N-2)r_+^{N-2}}\right)=\f{Q}{(N-2)r_+^{N-2}}.
\label{rnpot}
\end{eqnarray}
Now exploiting (\r{m}), (\r{q}), (\r{temp})and (\r{rnpot}) and using (\r{a}), (\r{an-3}) one gets the desired result for the Smarr formula, as given by
\begin{eqnarray}
M-Q\Phi=\f{N-1}{N-2}\f{\kappa A_{H}}{8\pi}.
\label{rnsmarr}
\end{eqnarray}
Note that for $Q=0$ it yields the correct result for the $N+1$ dimensional Schwarzschild black hole \cite{myers} and for $N=3$ it reduces to the well-known result for 3+1 dimensional R-N black hole \c{poisson}. 

Although the above algebraic approach successfully yields the Smarr formula for the R-N black holes, unfortunately this cannot be generalized to the rotating case. The reason for this is that we cannot write the horizon radius as a function of mass and angular momentum, mimicing (\r{rnhor}). A similar problem arises for the charged, rotating case (\r{metric}). In the subsequent part of this paper we develop a new technicque, using the concept of Killing symmetries and effective Komar conserved quantities, which solves this problem.

\section{Killing vectors, conserved charges and the identity $K_{\chi^{\mu}}=2ST$}

A spacetime encoded with symmetries is known to have conserved physical entities. These symmetries are actually characterised by the Killing vectors. Depending upon the number of these vectors one can associate an equal number of conserved quantities. These are known as Komar conserved quantities \cite{komar,komar2}. For an axially symmetric stationary spacetime one has multiple Killing vectors. In fact for the metric (\r{metric}) there are two Killing vectors $\xi^{\mu}_{(t)}=(1,0,0,0\cdot\cdot\cdot)$ and $\xi^{\mu}_{(\ph)}=(0,0,0,1,\cdot\cdot\cdot)$. The conserved quantities corresponding to these vectors are given by \c{komar,komar2}
\begin{eqnarray}
K_{\xi^{\mu}_{(t)}}=-\f{1}{8\pi G}\int \xi^{\mu;\nu}_{(t)}d^{N-1}\S_{\mu\nu}
\label{meff}
\end{eqnarray}
and
\begin{eqnarray}
K_{\xi^{\mu}_{(\ph)}}=-\frac{1}{8\pi G}\int \xi^{\mu;\nu}_{(\ph)}d^{N-1}\S_{\mu\nu}.
\label{jeff}
\end{eqnarray}
respectively. 

When these expressions are calculated at asymptotic infinity they give distinct black hole parameters (mass or angular momentum) upto some normalisation constant. For example take the case of (3+1) dimensional Kerr-Newman black hole where the results at asymptotic infinity are $\displaystyle\lim_{r\rightarrow\infty}K_{\xi^{\mu}_{(t)}}=M$ (the normalisation $(8\pi G)^{-1}$ is choosen so that the result matches with the Newtonian mass in the weak field approximation) and $\displaystyle\lim_{r\rightarrow\infty}K_{\xi^{\mu}_{(\ph)}}=-2J$. Based on these results one can now define a Komar mass ($M$) and Komar angular momentum ($J$) for (3+1) dimensional Kerr-Newman black hole in the following way
\begin{eqnarray}
M_{(3+1)}=-\f{1}{8\pi G}\displaystyle\int_{\Sigma(r\rightarrow\infty)} \xi^{\mu;\nu}_{(t)}d^{2}\S_{\mu\nu}
\label{meffkn}
\end{eqnarray}
and
\begin{eqnarray}
J_{(3+1)}=\frac{1}{16\pi G}\displaystyle\int_{\Sigma(r\rightarrow\infty)} \xi^{\mu;\nu}_{(\ph)}d^{2}\S_{\mu\nu}.
\label{jeffkn}
\end{eqnarray}
These definitions are often found in text books\c{carrol}. However, note that while the normalisations between (\r{meff}) and (\r{meffkn}) do not differ they are not same for (\ref{jeff}) and (\r{jeffkn}). This difference is required for the correct identification of the black hole angular momentum ($J$). Incidentally, when we are dealing with higher dimensions like the spacetime metric (\r{metric}) the normalisation for the black hole mass (which should match with the Newtonian mass in the weak gravity limit) also needs to be changed from its usual value as appears in (\r{meff}). These identifications finally define the Komar mass and angular momentum for the $N+1$ dimensional charged Myers-Perry black hole \c{myers},
\begin{eqnarray}
M_{(N+1)}=-\f{(N-1)}{16\pi G(N-2)}\int_{\Sigma(r\rightarrow\infty)} \xi^{\mu;\nu}_{(t)}d^{N-1}\S_{\mu\nu}
\label{meffknh}
\end{eqnarray}
and
\begin{eqnarray}
J_{(N+1)}=\frac{1}{16\pi G}\int_{\Sigma(r\rightarrow\infty)} \xi^{\mu;\nu}_{(\ph)}d^{N-1}\S_{\mu\nu}.
\label{jeffknh}
\end{eqnarray}  
The importance of these subtle issues will be discussed in more detail later on.

Now we want to calculate the general expressions for the Komar integrals (\r{meff}) and (\r{jeff}). The motivation of doing that is to find the effective values $K_{\xi^{\mu}_{(t)}}$ and $K_{\xi^{\mu}_{(\ph)}}$ which are valid for the entire region, i.e. on and outside the event horizon. In order to do so the first step is to write these equations in the following coordinate free dual two form \c{cohen2,modak}, \c{wald}-\c{dadhich2}
\begin{eqnarray}
K_{\xi^{\mu}_{(t)}}=-\f{1}{8\pi G}\int {}^*d\s
\label{meff1}
\end{eqnarray}
\begin{eqnarray}
K_{\xi^{\mu}_{(\ph)}}=-\frac{1}{8\pi G}\int {}^*d\eta
\label{jeff1}
\end{eqnarray}
where the timelike and the spacelike one forms for the spacetime metric (\r{metric}) are respectively given by
 $\s =\xi_{(t)\mu}dx^{\mu}=g_{0\m}dx^{\m}=g_{00}~dt+g_{03}~d\ph$,      $~~\eta=\xi_{(\ph)\mu}dx^{\mu}=g_{3\m}dx^{\m}=g_{30}~dt+g_{33}~d\ph$.
This yields the following two forms
\begin{eqnarray}
d\s &=&\p_rg_{00}~dr\wedge dt+\p_{\theta}g_{00}~d\th \wedge dt+\p_rg_{03}~dr\wedge d\ph+\p_{\th}g_{03}~d\th\wedge d\ph
\label{r1}\\
d\eta &=&\p_rg_{03}~dr\wedge dt+\p_{\theta}g_{03}~d\th \wedge dt+\p_rg_{33}~dr\wedge d\ph+\p_{\th}g_{33}~d\th\wedge d\ph
\label{tform}
\end{eqnarray}
Now calculating the Hodge duals of (\r{r1}) and (\r{tform}), we write the integrals (\r{meff1}) and (\r{jeff1})(see appendix B for details) as, 
\begin{eqnarray}
K_{\xi^{\mu}_{(t)}} &=&-\f{1}{8\pi G}\int r^{N-3}\cos^{N-3}\th\Lambda(r,\th)d\th d\ph d\chi_1\cdot\cdot\cdot\sin\chi_{N-4}d\chi_{N-3}\label{meff2}\\
K_{\xi^{\mu}_{(\ph)}} &=& -\f{1}{8\pi G}\int r^{N-3}\cos^{N-3}\th\Psi(r,\th)d\th d\ph d\chi_1\cdot\cdot\cdot\sin\chi_{N-4}d\chi_{N-3}\label{jeff3}
\end{eqnarray}
where $\Lambda$, $\Psi$ are defined as
\begin{eqnarray}
\Lambda(r,\th) &=&\sqrt{g_{22}g_{33}}\Lambda_{10}=\left(g_{33}\p_r g_{00}-g_{03}\p_r g_{03}\right)\sqrt{\f{g_{22}}{g_{33}(-g_{00}g_{33}+g_{03}^2)}},
\label{Lr}\\
\Psi(r,\th) &=&-\left(g_{33}\p_r g_{03}-g_{03}\p_r g_{33}\right)\sqrt{\f{g_{22}}{g_{11}(g_{03}^2-g_{00}g_{33})}}
\end{eqnarray}
To find the effective values of $K_{\xi^{\mu}_{(t)}}$ and $K_{\xi^{\mu}_{(\ph)}}$ the above integrations (\ref{meff2}, \r{jeff3}) need to be performed over all the angular variables ($\th,~\phi,~\chi_1,\cdot\cdot\chi_{N-3}$). 

Let us first separate different parts of the integral (\r{meff2}) and (\r{jeff3}) in the following manner 
\begin{eqnarray}
K_{\xi^{\mu}_{(t)}}=\frac{1}{8\pi G}\int_{\ph =0}^{2\pi}d\ph ~\int_{\th =0}^{\pi}{r^{(N-3)}\cos^{N-3}\th}~\Lambda(r,\th)d\th\nonumber\\
\int_{\chi_1,\chi_2,\cdot\cdot\cdot,\chi_{N-3} =0}^{\pi}~d\chi_1 \sin\chi_1\cdot\cdot\cdot\sin{\chi_{N-4}}~d\chi_{N-3},
\label{meff3}\\
K_{\xi^{\mu}_{(\ph)}}=\frac{1}{8\pi G}\int_{\ph =0}^{2\pi}d\ph ~\int_{\th =0}^{\pi}{r^{(N-3)}\cos^{N-3}\th}~\Psi(r,\th)d\th\nonumber\\
\int_{\chi_1,\chi_2,\cdot\cdot\cdot,\chi_{N-3}=0}^{\pi}~d\chi_1 \sin\chi_1\cdot\cdot\cdot\sin{\chi_{N-4}}~d\chi_{N-3}.
\end{eqnarray}
Performing the integrations over the azimuthal angle ($\ph$) and angular variables ($\chi_1,\chi_2,\cdot\cdot\cdot\chi_{N-3}$), we find
\begin{eqnarray}
K_{\xi^{\mu}_{(t)}}=\frac{\bar{A}_{N-3}}{4G}~\int_{\th =0}^{\pi}{r^{(N-3)}\cos^{N-3}\th}~\Lambda(r,\th)d\th
\label{meff4}\\
K_{\xi^{\mu}_{(\ph)}}=\frac{\bar{A}_{N-3}}{4G}~\int_{\th =0}^{\pi}{r^{(N-3)}\cos^{N-3}\th}~\Psi(r,\th)d\th.
\end{eqnarray}
where $\bar{A}_{N-3}$ has been defined in (\r{an-3}). Now an explicit integration over the polar angle ($\th$) gives the desired results for the effective values of different Komar conserved quantities for arbitrary dimensional charged Myers-Perry black holes. These are given by 
\begin{eqnarray}
K_{\xi^{\mu}_{(t)}}=\frac{\bar{A}_{N-3}}{G}\left(\frac{m}{2}-\frac{q^2}{r^{N-2}}-\frac{q^2a^2(a^2+r^2)}{Nr^{N+2}}~2F_{1}(2,\frac{N}{2};\frac{N}{2}+1;-\frac{a^2}{r^2})\right),
\label{meff5}
\end{eqnarray}
and
\begin{eqnarray}
K_{\xi^{\mu}_{(\ph)}} &=& -\frac{a\bar{A}_{N-3}}{4Gr^{N+4}}(K_1-K_2)\label{jeff5}\\
{\textrm{where,}}\nonumber\\
K_{1} &=& \frac{4mNr^{N+4}+2q^2r^2[a^4(N-2)^2+a^2(N-4)(N-2)r^2-4(N-1)r^4]}{N(N-2)}\\
K_{2} &=& \frac{2a^2(N-2)q^2(a^2+r^2)^2~ 2F_{1}(1,\frac{N+2}{2};\frac{N+4}{2};-\frac{a^2}{r^2})}{N+2}.
\end{eqnarray}
In the above expressions $2F_{1}$ is a hypergeometric function,
\begin{eqnarray}
2F_1(b,c;d;-\f{a^2}{r^2})=1+\f{b.c}{1!d}(-\f{a^2}{r^2})+\f{b(b+1).c(c+1)}{2!d(d+1)}(-\f{a^2}{r^2})^2+\cdot\cdot\cdot .
\end{eqnarray}
 Note that for a finite $r$, the effective values of $K_{\xi^{\mu}_{(t)}}$ and $K_{\xi^{\mu}_{(\ph)}}$ differ from their asymptotic values only due to presence of electric charge ($q$) in black holes. The extra contributions come in two different ways, one is proportional to the electric charge and the other is a coupling between the charge and the reduced angular momentum parameter ($a$). The second one can be termed as a gravito-electric effect. In the asymptotic limit all contributions due to the electric charge drop out and we find,  
\begin{eqnarray}
\lim_{r\rightarrow\infty}~K_{\xi^{\mu}_{(t)}} &=&\frac{m\bar{A}_{N-3}}{2G}\label{asykt}\\ \lim_{r\rightarrow\infty}~K_{\xi^{\mu}_{(\ph)}} &=&-\frac{ma\bar{A}_{N-3}}{(N-2)G}.
\label{asymkp}
\end{eqnarray}
Comparing these relations with (\r{m}), (\r{j}) and using the relations (\r{a}), (\r{an-3}) one finds,
\begin{eqnarray}
\lim_{r\rightarrow\infty}~K_{\xi^{\mu}_{(t)}} &=& \frac{2(N-2)}{(N-1)}M\label{asykt1}\\ \lim_{r\rightarrow\infty}~K_{\xi^{\mu}_{(\ph)}} &=& -2J\label{asmkp1}.
\label{compr}
\end{eqnarray}
From these two equations it is now clear that only in (3+1) dimensions the asymptotic value of the Komar conserved quantity, corresponding to the Killing vector $\xi^{\mu}_{(t)}=(1,0,0,0)$, gives the correct value of the black hole mass ($M$). For any other higher dimension the value of $K_{\xi^{\mu}_{(t)}}$ differs from the black hole mass by a  dimension dependent numerical factor. The other conserved quantity $K_{\xi^{\mu}_{(\ph)}}$ differs by $-2$ factor from the angular momentum ($J$) for all spacetime dimensions greater than or equal to four.

So far we have shown a complete scheme for calculating two Komar conserved quantities without using the asymptotic approximation. The motivation of doing that is to find the value of the Komar conserved quantity at the black hole event horizon. At the event horizon none of $\xi^{\mu}_{(t)}$ or $\xi^{\mu}_{(\ph)}$ are Killing vectors. To see this recall that for a vector $\chi^{\mu}$ to be Killing at the event horizon one must have $\chi^{\mu}\chi_{\mu}|_{r=r_+}=0$, i.e. it should be null at the event horizon. However for the spacetime metric (\r{metric}) none of the above two vectors satisfy this condition. Only a specific linear combination of those vectors, given by, $\chi^\mu=\xi^{\mu}_{(t)}+\Omega_{H}\xi^{\mu}_{(\ph)}$ (where $\Omega_H$ is the angular velocity at the event horizon) satisfies this condition. It can be verified that the use of $\chi^{\mu}$ leads to the correct result for the surface gravity $\kappa=\sqrt{-\f{1}{2}(\nabla^{\mu}\chi^{\nu})(\nabla_{\mu}\chi_{\nu})}\big|_{r=r_+}=\f{F'{(r_+)}}{2}$ (\r{surface}) which is related to the Hawking temperature (\ref{temp}) through the relation $T=\f{\kappa}{2\pi}$. Thus the black hole event horizon is a Killing horizon of the Killing vector $\chi^{\mu}$. Since this is a Killing vector one can again associate a Komar conserved quantity at the event horizon, which can be finally brought to the following form   
\begin{eqnarray}
K_{\chi^{\mu}}=K_{\xi^{\mu}_{(t)}}+\Omega_{H}~K_{\xi^{\mu}_{(\ph)}}
\label{kev}
\end{eqnarray}
Now it is easy to calculate this quantity by using the horizon condition $\Delta|_{r_+}=0$ and by using equations (\r{ang}), (\r{meff5}) and (\r{jeff5}). This result is found to be
\begin{eqnarray}
K_{\chi^{\mu}}=2 \left(\f{\pi r_+^{N-3}(r_+^2+a^2)\bar{A}_{N-3}}{N-2}\right)\left(\f{(N-4)(r_+^2+a^2)+2r_+^2-(N-2)q^2r_+^{2(3-N)}}{4\pi r_+(r_+^2+a^2)}\right)
\label{lhs}
\end{eqnarray}
where we have used the following identities involving the hypergeometric functions, 
\begin{eqnarray}
&&\f{Na^2}{(N+2)r^2}2F_{1}(1,\f{N+2}{2};\f{N+4}{2};-\f{a^2}{r^2})+2F_{1}(1,\f{N}{2};\f{N+2}{2};-\f{a^2}{r^2})=1
\label{iden1}\\
&&2F_{1}(2,\f{N}{2};\f{N+2}{2};-\f{a^2}{r^2})=\f{N}{2(1+\f{a^2}{r^2})}+(1-\f{N}{2})2F_{1}(1,\f{N}{2};\f{N+2}{2};-\f{a^2}{r^2})
\label{iden2}
\end{eqnarray}
It is very interesting to note that, using (\r{entr}) and (\r{temp}), equation (\r{lhs}) can be written as 
\begin{eqnarray}
K_{\chi^{\mu}}=2ST,
\label{2st}
\end{eqnarray}
i.e. the Komar conserved quantity at the event horizon corresponding to the null Killing vector is twice the product of Hawking temperature and black hole entropy. A remarkable feature of the above relation is that it is independent of the number of dimensions $N$.

It is possible to make a correspondence of (\r{2st}) with a recent relation,
\begin{eqnarray}
E=2ST,
\label{e=2st}
\end{eqnarray}
discussed in the literature \c{modak},\c{paddy1}-\c{majhi}. Here $E$ is the conserved Noether charge corresponding to a diffeomorphic transformation that exists with or without a black hole metric. In the case of a black hole we are able to precisely identify $E$ as the conserved Noether charge that is equal to a particular combination of properly normalised Komar conserved quantities given by (\ref{kev}). For the particular case of $3+1$ dimensions only, the relation (\r{2st}) was found earlier by two of us \c{modak}. 

\section{Smarr formula for charged Myers-Perry black holes}
 To understand the significance of (\r{2st}) in arbitrary dimension we proceed as follows. First note that, the right  hand side of (\r{2st}) can be written in terms of surface gravity ($\kappa$) and horizon area ($A_H$) of the black hole as $2(\f{A_H}{4})(\f{\kappa}{2\pi})$. Now we want to rewrite the left hand side of (\r{2st}) as a combination of three seperate terms, one invoving $m$, one involving $ma$ and the other involving $q^2$. To do that we use (\r{kev}), (\r{meff5}), (\r{jeff5}) together with the identities (\r{iden1}), (\r{iden2}) and the expression of $\Omega_H$ (\r{ang}). This gives
\begin{eqnarray}
\f{\bar{A}_{N-3}}{G}\left(\f{m}{2}-\f{ma\Omega_H}{N-2}-\f{q^2r_+^{2-N}[a^2(N-3)+r_+^2(N-2)]}{(N-2)(a^2+r_+^2)}\right)=2\f{\kappa A_H}{8\pi}
\label{}
\end{eqnarray}
Making use of the relations (\r{m}), (\r{j}) and (\r{q}) we write the above relation as,
\begin{eqnarray}
\f{2M(N-2)}{(N-1)}-2\Omega_H J-\f{2Q^2r_+^{4-N}}{(N-1)(a^2+r_+^2)}-\f{2Q^2r_+^{2-N}a^2(N-3)}{(N-1)(N-2)(a^2+r_+^2)}=2\f{\kappa A_H}{8\pi}
\label{smarr}
\end{eqnarray}

The scalar potential $\Phi$ at the event horizon is given by the timelike component of (\r{phi}). This is found by contracting $A_{\mu}$ with the Killing field, $\chi^{\mu}=\xi^{\mu}_{(t)}+\Omega_H\xi^{\mu}_{(\ph)}=(1,0,0,\Omega_H,0,\cdot\cdot\cdot)$ corresponding to the metric (\r{metric}) and is given by
\begin{eqnarray}
\Phi=\chi^{\mu}A_{\mu}\big|_{r=\infty}-\chi^{\mu}A_{\mu}\big|_{r=r_+}=\f{Q}{(N-2)r_+^{N-4}(r_+^2+a^2)}
\label{ph}
\end{eqnarray}
Using the above relation, (\r{smarr}) is written as,
\begin{eqnarray}
\f{M(N-2)}{(N-1)}-\Omega_H J-Q\Phi\left(\f{(N-2)}{N-1}+\f{a^2(N-3)}{r_+^2(N-1)}\right)=\f{\kappa A_H}{8\pi}
\label{sm1}
\end{eqnarray}
Structurally this is very similar to the Smarr formula. However there is only one unfamiliar term (involving $a^2$) arising at the left hand side of (\r{sm1}). Let us now recall that the spacetime (\ref{metric}) is a solution of the Einstein-Maxwell equation only for slowly rotating case (upto linear order in $a$) \cite{aliev}. Therefore we can drop this unfamiliar term and rewrite (\r{sm1}) in the following way, 
\begin{eqnarray}
M(N-2)=(N-1)\Omega_H J+(N-2)Q\Phi +(N-1)\f{\kappa A_H}{8\pi}.
\label{sm2}
\end{eqnarray}
This is the cherished Smarr formula for the $N+1$ dimensional charged Myers-Perry black hole. In the following we show that (\r{sm2}) is indeed compatible with the first law of black hole thermodynamics.

Looking at the metric (\r{metric}) we can express the following three basic variables in terms of their dimensionless {\it primed} counterparts, given by
\begin{eqnarray}
m=\l^{N-2} m',~a=\l a',~q=\l^{N-2} q'~,
\label{sm3}
\end{eqnarray}
where $\l$ is a dimensionful quantity having the dimension of length. Now using (\r{sm3}) it is easy to check that the horizon radius can be expressed as $r_+=\l r_+'$. Exploiting these four basic transformations in (\r{m}), (\r{j}), (\r{q}) and (\r{hora1}) we obtain the following relations
\begin{eqnarray}
M=\l^{N-2} M',~J=\l^{N-1}J',~Q=\l^{N-2} Q',~A_{H}=\l^{N-1}A_{H}'~ .
\label{sm4}
\end{eqnarray}
Now if we treat $M$, $J$, $A_{H}$ and $Q$ as independent variables then (\r{sm2}) is a homogeneous equation of degree $(N-2)$ in $M$, $(N-1)$ in $J$, $(N-2)$ in $Q$ and $(N-1)$ in $A_H$. Therefore one can now exploit Euler's theorem on homegeneous functions to extract the differential form of Smarr formula,
\begin{eqnarray}
dM=\f{\kappa}{8\pi}~dA_{H}+\Omega_{H} ~dJ+\Phi ~dQ.
\end{eqnarray}   
This is the desired result of the first law of black hole thermodynamics for rotating, charged black holes in arbitrary dimensions, compatible with (\r{sm2}).

We now provide certain consistency checks on this new Smarr formula (\r{sm2}). 
In 3+1 dimensions 
the resulting relation is the Smarr formula for the Kerr-Newman black hole. This was originally given by Smarr \c{smarr} and also by Bardeen, Carter and Hawking \c{Bardeen}. 
Moreover (\r{sm2}) is consistent with the result for the R-N black hole as derived in (\r{rnsmarr}). Finally, this result also gives the correct Smarr formula for the arbitrary dimensional (charge-less) Myers-Perry black hole\c{myers}.

\section{Conclusions}
Let us summarize the main results of our paper. We have calculated the semiclassical Hawking temperature ($T$) and the entropy ($S$) of an arbitrary dimensional charged Myers-Perry black hole. Then we introduced a formulation where the Komar conserved charges corresponding to appropriate Killing vectors were calculated at any arbitrary point on or outside the event horizon. This may be compared with conventional mathods where these conserved charges are calculated either at the horizon or at infinity. Using these results, we found the conserved quantity ($K_{\chi^{\mu}}$) at the event horizon for the null Killing vector ($\chi^{\mu}$) which was a combination of two vectors which are, respectively,  asymptotically time-like and space-like. Interestingly, $K_{\chi^{\mu}}$ satisfies the remarkable relation $K_{\chi^{\mu}}=2ST$ for any dimensions.

The striking nature of $K_{\chi^{\mu}}=2ST$ was that in spite of the explicit dimensional dependence of all individual quantities ($K_{\chi^{\mu}}$ (\r{lhs}), $S$ (\r{entr}), $T$ (\r{temp})) the relation itself was completely dimension independent. Furthermore, no explicit black hole parameters (like mass, angular momentum or charge) appeared in this relation. This indicates the possibility of the identity being valid in general and not confined to just black holes. In fact, based on a Lagrangian approach, Padmanabhan \c{paddy1}-\c{paddy4} suggested the possibility of such an identity whenever a local Killing horizon existed. In that case the left hand side of the identity was a conserved Noether charge corresponding to a diffeomorphic transformation and was represented by a new improved surface integral. The explicit demonstration of this identity as well as the definition of the surface integral in terms of Komar integrals in the black hole context was a new result of our paper. As an application of this identity we also derived the generalized Smarr formula for charged Myers-Perry black holes in any dimension. Several consistency checks were performed which reassured the validity of this new formula. Using the Euler's theorem on homogeneous functions the compatibility of the generalised Smarr formula with the first law of black hole thermodynamics was demonstrated. In fact we also exploited this theorem to generalise the original work of Smarr \c{smarr} to obtain the Smarr formula for arbitrary dimensional Reissner-Nordstrom black holes. However this approach could not be generalised to the rotating case.

Regarding future prospects we cite a recent work by Barnich and Compere \c{barnich} who provided a new approach to derive the generalised Smarr formula for higher dimensional rotating black holes in flat or AdS background. From this approach it might be possible to find the above dimension independent identity (\r{2st}) where the improved surface integrals could be used to define the left hand side. This optimism is borne by the fact that notions of improved surface integrals were also used in the Lagrangian approach \c{paddy1}-\c{paddy4} to define the identity. Such surface integrals are expected to provide a connection between the Lagrangian and Hamiltonian approaches. We leave these and related issues for a future publication.

\section*{Appendix A: Dimensional Reduction for Charged Myers--Perry Black Hole}
\renewcommand{\theequation}{A.\arabic{equation}}
\setcounter{equation}{0}
        We consider the complex scalar action under the background metric (\ref{metric}):
\begin{eqnarray}
{\cal{A}} &=& -\int d^{N+1}X \varphi^* (\nabla_\mu + iA_\mu)(\nabla^\mu-iA^\mu)\varphi
\label{new1}
\end{eqnarray}
where $d^{N+1}X = \sqrt{-g} dt dr d\theta d\phi d\chi_1 ...... d\chi_{N-3}$ with $\sqrt{-g}$ given by (\ref{deter}). Now substituting the expansion for $\varphi$ in terms of spherical harmonics
\begin{eqnarray}
\varphi = \sum_{m_1,m_2,...,m_{N-1}}\varphi_{m_1,m_2,...,m_{N-1}}(t,r)Y_{m_1,m_2,...,m_{N-1}}(\theta,\phi,\chi_1,....,\chi_{N-3})
\label{new2}
\end{eqnarray}
and then using 
\begin{eqnarray}
&&dr^* = \frac{(r^2+a^2)r^{N-2}}{\Delta}dr
\label{dim12}
\nonumber
\\
&&\hat{L}_\phi Y_{m_1,m_2,...,m_{N-1}}(\theta,\phi,\chi_1,....,\chi_{N-3}) = m_2 Y_{m_1,m_2,...,m_{N-1}}(\theta,\phi,\chi_1,....,\chi_{N-3})
\label{dim2}
\end{eqnarray}
we obtain,
\begin{eqnarray}
{\cal{A}}&=& \sum_{m'_1,m'_2,...,m'_{N-1}}\varphi^*_{m'_1,m'_2,...,m'_{N-1}}(t,r)Y^*_{m'_1,m'_2,...,m'_{N-1}}(\theta,\phi,\chi_1,....,\chi_{N-3})\nonumber\\&&\int dt~dr^* d\Omega_{N-1}\sqrt{\gamma}r^{N-3}~\sin\theta~\cos^{N-3}\theta 
\Big[(r^2+a^2) (\partial_t+iA_t)^2
- \partial_{r^*}(r^2+a^2)\partial_{r^*} 
\nonumber
\\
&+& \frac{\Delta}{(r^2+a^2)r^{N-2}}({\hat G} + {\hat L}^2_{(\theta,\chi_1,....,\chi_{N-3})})\Big]\nonumber\\&&\sum_{m_1,m_2,...,m_{N-1}}\varphi_{m_1,m_2,...,m_{N-1}}(t,r)Y_{m_1,m_2,...,m_{N-1}}(\theta,\phi,\chi_1,....,\chi_{N-3})
\label{dim1.01}
\end{eqnarray}
where,
\begin{eqnarray}
A_t = -eV(r) - m_2\Omega(r); V(r) = \frac{Q}{(N-2)r^{N-4}(r_+^2+a^2)}; \Omega = \frac{a}{r^2+a^2}
\label{dim3}
\end{eqnarray}
and ${\hat G}$,  ${\hat L}^2_{(\theta,\chi_1,....,\chi_{N-3})}$ are operators whose explicit forms are not necessary and $e$ is the charge of the complex scalar field. Since near the horizon $\Delta\rightarrow 0$, the scalar action reduces to
\begin{eqnarray}
{\cal{A}} &\simeq& \sum_{m'_1,m'_2,...,m'_{N-1}}\varphi^*_{m'_1,m'_2,...,m'_{N-1}}(t,r)Y^*_{m'_1,m'_2,...,m'_{N-1}}(\theta,\phi,\chi_1,....,\chi_{N-3})\nonumber\\&&\int dt~dr^*d\Omega_{N-1} \sqrt{\gamma}r^{N-3}~\sin\theta~\cos^{N-3}\theta 
\Big[(r^2+a^2) (\partial_t+iA_t)^2
- \partial_{r^*}(r^2+a^2)\partial_{r^*} 
\Big]\nonumber\\&&\sum_{m_1,m_2,...,m_{N-1}}\varphi_{m_1,m_2,...,m_{N-1}}(t,r)Y_{m_1,m_2,...,m_{N-1}}(\theta,\phi,\chi_1,....,\chi_{N-3})
\label{dim1.02}
\end{eqnarray}
Now using the orthonormal condition 
\begin{eqnarray}
&&\int d\theta d\phi d\chi_1 .... d\chi_{N-3} \sin\theta \cos^{N-3}\theta \sin\chi_1 .... \sin\chi_{N-4}
\nonumber
\\
&&Y^*_{m_1,m_2,...,m_{N-1}}(\theta,\phi,\chi_1,....,\chi_{N-3}) Y_{m'_1,m'_2,...,m'_{N-1}}(\theta,\phi,\chi_1,....,\chi_{N-3})
\nonumber
\\
&=&\delta_{m'_1,m_1}.......\delta_{m'_{N-1},m_{N-1}}
\label{new3}
\end{eqnarray}
we obtain,
\begin{eqnarray}
{\cal{A}} &\simeq & \sum_{m_1,m_2,...,m_{N-1}}\varphi^*_{m_1,m_2,...,m_{N-1}}(t,r)\int dt~dr^* r^{N-3}
\nonumber
\\
&&\Big[(r^2+a^2) (\partial_t+iA_t)^2
- \partial_{r^*}(r^2+a^2)\partial_{r^*} 
\Big]\varphi_{m_1,m_2,...,m_{N-1}}(t,r)
\label{dim1.03}
\end{eqnarray}
Reverting back to the original $r$ coordinate,
\begin{eqnarray}
{\cal{A}} &\simeq & 
-\sum_{m_1,m_2,....,m_{N-1}} \int dt~dr H(r_+) \varphi^*_{m_1,m_2,....,m_{N-1}}(t,r)\Big[-\frac{(r^2+a^2)r^{N-2}}{\Delta} (\partial_t+iA_t)^2 
\nonumber
\\
&+& \partial_r\frac{\Delta}{(r^2+a^2)r^{N-2}}\partial_r \Big]\varphi_{m_1,m_2,....,m_{N-1}}(t,r)
\label{dim4}
\end{eqnarray}
Here $H$ is a function of $r_+$ whose explicit expression is not required for our analysis.
This shows that each partial wave mode of the fields can be described near the horizon as a (1 + 1) dimensional complex scalar field with two U(1) gauge potentials $V (r)$,$\Omega(r)$ and the dilaton field $\psi = H(r_+)$. It should be noted that the above action for each $m_1,m_2,....,m_{N-1}$ can also be obtained from a complex scalar field action in the background of metric 
\begin{eqnarray}
ds^2_{eff} = -F(r)dt^2 + \frac{dr^2}{F(r)},
\label{effective1}
\end{eqnarray}
where
\begin{eqnarray}
F(r) = 1 - \frac{m}{r^{N-4}(r^2+a^2)} + \frac{q^2}{r^{2(N-3)}(r^2+a^2)}
\label{effective2}
\end{eqnarray}
with the dilaton field $\psi$. For detailed study on dimensional reduction see \cite{robin,kumet,iso1}. Subsequently we use this effective metric to derive the Hawking temperature (\ref{temp}).

\section*{Appendix B: Calculation of Komar Integrands}
\renewcommand{\theequation}{B.\arabic{equation}}
\setcounter{equation}{0}
We introduce the following orthonormal basis for the spacetime metric (\r{metric})
\begin{eqnarray}
\varkappa_0&=&-\sqrt{\left(\f{g_{03}^2}{g_{33}}-g_{00}\right)}~dt\nonumber\\
\varkappa_1&=&\sqrt{g_{11}}dr\nonumber\\
\varkappa_2&=&\sqrt{g_{22}}d\th\nonumber\\
\varkappa_3&=&\sqrt{g_{33}}(d\ph+\f{g_{03}}{g_{33}}dt)\label{vark}\\
\varkappa_4&=&r\cos\th d\chi_1\nonumber\\
\varkappa_5&=&r\cos\th\sin\chi_1 d\chi_2\nonumber\\
\varkappa_i&=&r\cos\th\sin\chi_1\cdot\cdot\cdot\sin\chi_{i-4}d\chi_{i-3} \  \ ({\textrm{where}} \ 6\le i\le N+1)\nonumber
\end{eqnarray}
in which the metric (\r{metric}) takes the Minkwoskian form
\begin{eqnarray}
ds^2=-\varkappa^2_0+\varkappa_1^2+\varkappa_2^2+\cdot\cdot\cdot+\varkappa_N^2
\label{metricvar}
\end{eqnarray}
Using the inverse relations,
\begin{eqnarray}
dt&=&-\f{1}{\sqrt{\f{g_{03}^2}{g_{33}}-g_{00}}}\varkappa_0\nonumber\\
dr&=&\f{1}{\sqrt{g_{11}}}\varkappa_1\nonumber\\
d\th &=&\f{1}{\sqrt{g_{22}}}\varkappa_2\nonumber\\
d\ph&=&\f{1}{\sqrt{g_{33}}}\left(\varkappa_3+\f{g_{03}}{\sqrt{g_{00}g_{33}+g_{03}^2}}\varkappa_0\right)\\
d\chi_1&=&(r\cos\th)^{-1} \varkappa_4\nonumber\\
d\chi_2&=&(r\cos\th\sin\chi_1)^{-1}\varkappa_5\nonumber\\
d\chi_{i-3}&=&(r\cos\th\sin\chi_1\cdot\cdot\cdot\sin\chi_{i-4})^{-1}\varkappa_i \  \ ({\textrm{where}} \ 6\le i\le N+1)\nonumber
\label{vark1}
\end{eqnarray}
(\r{r1}) is written as
\begin{eqnarray}
d\s=\Lambda_{10}\varkappa_1\wedge\varkappa_0+\Lambda_{20}\varkappa_2\wedge\varkappa_0
+\Lambda_{13}\varkappa_1\wedge\varkappa_3+\Lambda_{23}\varkappa_2\wedge\varkappa_3
\label{aaa}
\end{eqnarray}
where
\begin{eqnarray}
\Lambda_{10}&=&-\f{1}{\sqrt{g_{03}^2-g_{00}g_{33}}}\left(\sqrt{\f{g_{33}}{g_{11}}}\p_rg_{00}+\f{g_{03}}{\sqrt{g_{11}g_{33}}}\p_rg_{03}\right) \label{lam10} \\
\Lambda_{20}&=& -\f{1}{\sqrt{g_{03}^2-g_{00}g_{33}}}\left(\sqrt{\f{g_{33}}{g_{22}}}\p_{\th} g_{00}+\f{g_{03}}{\sqrt{g_{22}g_{33}}}\p_{\th}g_{03}\right) \\
\Lambda_{13}&=&\f{1}{\sqrt{g_{11}g_{33}}}\p_rg_{03}  \\
\Lambda_{23}&=& \f{1}{\sqrt{g_{22}g_{33}}}\p_{\th}g_{03} \\
\end{eqnarray}
The Hodge dual of (\r{aaa}) is
\begin{eqnarray}
{}^*d\s&=&\left(-\Lambda_{10}\varkappa_2\wedge\varkappa_3+\Lambda_{20}\varkappa_1\wedge\varkappa_3
+\Lambda_{13}\varkappa_2\wedge\varkappa_0-\Lambda_{23}\varkappa_1\wedge\varkappa_0\right)
\wedge\varkappa_4\wedge\varkappa_5\wedge\cdot\cdot\cdot
\end{eqnarray}
Using (\r{vark}), the above expression is written in the usual coordinates as,
\begin{eqnarray}
{}^*d\s&=&(-\Lambda_{10}\sqrt{g_{22}g_{33}}d\th\wedge d\ph+[-\Lambda_{10}g_{03}\sqrt{\f{g_{22}}{g_{33}}}-\Lambda_{13}\sqrt{\f{g_{22}}{g_{33}}}(g_{03}^2-g_{00}g_{33})^{\f{1}{2}}]d\th\wedge dt\nonumber\\
&&+\Lambda_{20}\sqrt{g_{11}g_{33}}dr\wedge d\ph+[\Lambda_{20}g_{03}\sqrt{\f{g_{11}}{g_{33}}}+\Lambda_{23}\sqrt{\f{g_{11}}{g_{33}}}(g_{03}^2-g_{00}g_{33})^{\f{1}{2}}]dr\wedge dt
)\wedge\nonumber\\&&(r\cos\th d\chi_1)\wedge(r\cos\th\sin\chi_1 d\chi_2)\wedge\cdot\cdot\cdot
\end{eqnarray}
At this point we need to choose an appropriate boundary ($\partial\Sigma$) characterised by a constant $r$ and $dt=-\frac{g_{03}}{g_{00}}d\phi$. This choice of surface reduces (\ref{meff1}) to
\begin{eqnarray}
K_{\xi^{\mu}_{(t)}}&=&\f{1}{8\pi G}\int \Lambda_{10}\sqrt{g_{22}g_{33}}(r\cos\th)^{N-3}
d\th d\ph d\chi_1\sin\chi_1 d\chi_2\cdot\cdot\cdot+\nonumber\\
&&\f{1}{8\pi G}\int\frac{g_{03}}{g_{00}}\left[\Lambda_{10}g_{03}\sqrt{\f{g_{22}}{g_{33}}}+\Lambda_{13}\sqrt{\f{g_{22}}{g_{33}}}(g_{03}^2-g_{00}g_{33})^{\f{1}{2}}\right](r\cos\th)^{N-3}d\th d\ph d\chi_1\sin\chi_1d\chi_2\cdot\cdot\cdot\nonumber
\end{eqnarray}
The second term of the right hand side measures the time shifting, when one moves along a closed contour. Since the calculation is perfomed over simultaneous events we subtract this term\cite{cohen1,cohen2} from the above integral to obtain,
\begin{eqnarray}
K_{\xi^{\mu}_{(t)}}&=&\f{1}{8\pi G}\int \Lambda_{10}\sqrt{g_{22}g_{33}}(r\cos\th)^{N-3}d\th d\ph d\chi_1\sin\chi_1d\chi_2\cdot\cdot\cdot
\end{eqnarray}
Making use of (\r{lam10}) we write the above equation as (\ref{meff2}). Similar consideration for the other Killing vector will give the relation (\r{jeff3}).

\section*{Acknowledgement} 
One of the authors (S.K.M) thanks the Council of Scientific and Industrial Research (C.S.I.R), Government of India, for financial support.


\end{document}